\documentclass[12pt]{article}

\usepackage[a4paper,left=30mm,right=20mm,top=20mm,bottom=20mm]{geometry}
\usepackage{graphics}
\usepackage{amsmath}
\usepackage{epsfig}
\usepackage{cite}

\title{On the possibility to detect the Higgs decay $H\to b\bar b$ in the associated $Z+ b\bar b$ production at the LHC}

\author{A.V.~Lipatov$^{1,\,2}$, N.P.~Zotov$^1$}

\begin{document}

\maketitle

\begin{center}

{\it $^1$Skobeltsyn Institute of Nuclear Physics, Lomonosov Moscow State University, 119991 Moscow, Russia}\\
{\it $^2$Joint Institute for Nuclear Research, Dubna 141980, Moscow Region, Russia}

\end{center}

\vspace{0.5cm}

\begin{center}

{\bf Abstract }

\end{center} 

\indent
We investigate the possibility to 
detect the scalar Higgs boson decay $H\to b\bar b$  
in the associated $Z$ and $b\bar b$ production at the LHC
using the $k_T$-factorization QCD approach.
Our consideration is based on the off-shell (i.e. depending on the 
transverse momenta of initial quarks and gluons) 
production amplitudes of $q^* \bar q^* \to Z H \to Z q^\prime \bar q^\prime$,
$q^* \bar q^* \to Z q^\prime \bar q^\prime$ and
$g^* g^* \to Z q^\prime \bar q^\prime$ 
partonic subprocesses supplemented with the Catani-Ciafoloni-Fiorani-Marchesini (CCFM)
dynamics of parton densities in a proton.
We argue that the $H \to b\bar b$ signal could be observed at large transverse 
momenta near Higgs boson peak despite the overwhelming QCD background, and point 
out an important role of angular correlations between the produced $Z$ 
boson and $b$-quarks.

\vspace{1.0cm}

\noindent
PACS number(s): 12.38.Bx, 14.65.Fy, 14.70.Hp, 14.80.Bn

\newpage
\indent

In 2012, during the search performed by the CMS and ATLAS Collaborations at the LHC,
the scalar Higgs boson $H$ with a mass $m_H$ near $125$ GeV has been discovered\cite{1,2},
giving us the confidence in the physical 
picture of fundamental interactions which follows from the Standard Model Lagrangian.
Some time later, the ATLAS Collaboration has reported first measurements
of the Higgs boson differential cross sections in the $\gamma \gamma$ decay mode\cite{3}.
The measured cross sections were found a bit higher than the central next-to-next-to-leading order 
(NNLO) expectations\cite{4,5,6,7,8,9} and those matched with soft-gluon resummation carried out up 
to next-to-next-to-leading logarithmic accuracy (NNLL)\cite{10,11}, although no significant
deviations from theoretical 
predictions are observed within the uncertainties\cite{3}.
The significant signal was detected also in channels where the Higgs boson
decays into the $ZZ$ or $WW$ pairs\cite{12,13}. The interaction of the Higgs
particle with the massive $Z$ and $W$ bosons indicates that, as it was expected, it plays 
a role in electroweak symmetry breaking. However, the interaction with the fermions and 
whether the Higgs field serves as the source of mass generation in the fermion sector 
still remains to be established. Since Higgs boson with mass $m_H \sim 125$~GeV decays 
mainly into a beauty quark-antiquark pair\cite{14}, the observation and study of the 
$H \to b\bar b$ decay (which involves the direct coupling of the Higgs boson to 
beauty quarks) is therefore essential in determining the nature of the newly discovered boson. 

The most sensitive channel for the $H \to b\bar b$ events at the LHC 
is the production of Higgs particle in association with the $Z$ boson\cite{15}.
Despite the largest branching fraction ($\sim 58$\%), 
the $H \to b \bar b$ final state is a more difficult for the
experimental observation compared to the  
signatures provided by the diphoton or diboson %$\gamma \gamma$, 
%$ZZ$ or $WW$ 
decay modes due to small signal over background ratio.
One of main backgrounds
for the associated Higgs and $Z$ boson production
is the associated production of $Z$ 
boson and two $b$-quark jets. 
The corresponding cross sections,
calculated at the NNLO level (see\cite{14}), are several orders of magnitude 
larger than the Higgs boson signal.
However, recently CMS Collaboration
reported\cite{16} an excess of events above the expected 
background with a local significance of $2.1$ standard
deviations, which is compatible with a Higgs boson mass of $125$ GeV. 
Earlier, the CDF and D0 Collaborations at the Tevatron also reported\cite{15} 
evidence for an excess of events in the $115 - 140$ GeV mass
range, consistent with the mass of the Higgs boson observed at the LHC. 

The experimental searches\cite{15,16} are stimulated us to 
investigate the associated Higgs (decaying into a $b\bar b$ pair) and $Z$ boson
production as well as corresponding main background process, associated production of $Z$ 
boson and two $b$-quark jets\footnote{Other background processes,
like as $t \bar t$ pair, diboson or 
QCD multijet production are out of our present consideration.},
using the $k_T$-factorization approach of QCD\cite{17,18}.
A detailed description of this
formalism can be found, for example, in reviews\cite{19}. We 
only mention that the main part of higher-order QCD corrections
(namely, NLO + NNLO + N$^3$LO + ... contributions which correspond to the $\log 1/x$ enhanced 
terms in perturbative series) is effectively taken into account in the $k_T$-factorization 
approach already at leading order, and 
it provides solid theoretical ground for the 
effects of initial parton radiation and transverse momenta of initial quarks and gluons.
Recently, the $k_T$-factorization QCD approach was successfully applied\cite{20,21} to 
describe the ATLAS data\cite{3} on 
the inclusive Higgs production
in the diphoton decay mode\footnote{In our opinion, the
results\cite{21} suffer from the problem of double counting and 
contain the wrong numerical factor.}. 

Let us start from a short review of calculation steps.
Our consideration is based on the off-shell (depending on the 
transverse momenta of initial partons) 
partonic subprocesses:
\begin{equation}
 q^*(k_1) + \bar q^*(k_2) \to Z + H \to Z (p) + q^\prime(p_1) + \bar q^\prime(p_2),
\end{equation}
\begin{equation}
  q^*(k_1) + \bar q^*(k_2) \to Z (p) + q^\prime(p_1) + \bar q^\prime(p_2),
\end{equation}
\begin{equation}
  g^*(k_1) + g^*(k_2) \to Z (p) + q^\prime(p_1) + \bar q^\prime(p_2),
\end{equation}

\noindent
where the four-momenta of all corresponding particles are given in
the parentheses (see Fig.~1). %Last two of these subprocesses correspond to the main QCD background
The subprocesses~(2) and (3) correspond to the main QCD background
to the associated Higgs and $Z$ boson production. 
Note, to calculate the production amplitudes, we apply the reggeized parton approach\cite{22,23}, 
which is based on the effective action formalism\cite{24},
currently explored at next-to-leading order\cite{25}, and
take into account the virtualities of both initial quarks and gluons.
In this point our consideration differs from the one based on 
the collinear QCD factorization, where these virtualities are not taken into account. 
The use of effective vertices\cite{22,23} ensures the exact gauge 
invariance of calculated amplitudes despite the off-shell initial partons.

The off-shell amplitude of 
subprocess~(1) reads:
\begin{equation}
  \displaystyle {\cal M}_1 = ee_q \, \epsilon^\mu(p)\, \bar v_{s_1}(p_2) \Gamma_{Hq\bar q} u_{s_{2}}(p_1) \, {1 \over (p_1 + p_2)^2 - m_H^2 - i m_H \Gamma_H } \, \times \atop {
    \displaystyle \times \, \Gamma^{\mu \nu}_{ZZH} \left[ g^{\nu \lambda} - { {(k_1 + k_2)^\nu(k_1 + k_2)^\lambda}\over m_Z^2}\right] {1\over \hat s - m_Z^2 - im_Z\Gamma_Z} \, \bar v_{r_1}(x_2 l_2) \Gamma^\lambda_{q^*\bar q^*Z} u_{r_2}(x_1 l_1) },
\end{equation}

\noindent 
where $e$ and $e_q$ are the electron and incoming quark (fractional) electric charges, 
$\epsilon^\mu$ is the polarization $4$-vector of produced $Z$ boson,
$\hat s = (k_1 + k_2)^2$, $k_i = x_i l_i + k_{i T}$ (with $i = 1$ or $2$),
$l_1$ and $l_2$ are the $4$-momenta of colliding protons,
$x_1$ and $x_2$ are the corresponding momentum fractions and
$\Gamma_H$ is the full decay width of Higgs boson, $m_Z$ and $\Gamma_Z$ 
are the mass and full decay width of $Z$ boson. 
We will take the propagators of intermediate $Z$ and Higgs bosons 
in the Breit-Wigner form to avoid any artificial singularities in the numerical calculations.
The fermion and gauge boson to Higgs vertices 
are usual:
\begin{equation}
 \Gamma_{Hq\bar q} = - {e \over \sin 2\theta_W} {m_{q^\prime}\over m_Z},
\end{equation}
\begin{equation}
 \Gamma_{ZZH}^{\mu \nu} = {e \over \sin 2\theta_W} \, g^{\mu \nu}m_Z ,
\end{equation}

\noindent 
where $m_{q^\prime}$ is the mass of produced quark or antiquark, $\theta_W$ is the Weinberg mixing angle.
We will neglect the masses of initial quarks compared to the masses of final state particles 
but keep their non-zero transverse momenta: ${\mathbf k}_{1T}^2 = - k_{1T}^2 \neq 0$,
${\mathbf k}_{2T}^2 = - k_{2T}^2 \neq 0$. The effective vertex
$\Gamma^\mu_{q^*\bar q^*Z}$ which describes the effective
coupling of off-shell quark and antiquark to $Z$ boson reads\cite{22,23} (see also\cite{26}):
\begin{equation}
  \Gamma^\mu_{q^*\bar q^*Z} = \left[\gamma^\mu - \hat k_1 {l_1^\mu \over (l_1 \cdot k_2)} - \hat k_2 {l_2^\mu \over (l_2 \cdot k_1)} \right] \left(C_V^q - C^q_A \gamma^5 \right),
\end{equation}

\noindent 
where $C_V^q$ and $C_A^q$
are the corresponding vector and axial coupling constants. The effective vertex
$\Gamma^\mu_{q^*\bar q^*Z}$ satisfy the Ward identity
$\Gamma^\mu_{q^*\bar q^*Z} (k_1 + k_2)_\mu = 0$. 
%It is obvious that the amplitude~(4) is gauge invariant despite the off-shell initial quarks.
The off-shell amplitude of subprocess~(2) reads:
\begin{equation}
  \displaystyle {\cal M}_2 = e e_{q^\prime} g^2 t^a \delta^{ab} t^b \epsilon^{\mu}(p) \, \bar v_{s_1}(p_2) F_{1}^{\mu \nu} u_{s_2}(p_1) \, {g^{\nu \lambda}\over (k_1 + k_2)^2} \, \bar v_{r_1}(x_2 l_2) \Gamma^\lambda_{q^*\bar q^* g} u_{r_2}(x_1 l_1) \, + \atop {
  \displaystyle + \, e e_q \, g^2 t^a \delta^{ab} t^b \epsilon^{\mu}(p) \, \bar v_{s_1}(p_2) F_{2}^{\mu \lambda} u_{s_2}(p_1) \, {g^{\nu \lambda} \over (p_1 + k_2)^2} \bar v_{r_1}(x_2 l_2) \gamma^\nu u_{r_2}(x_1 l_1) }, 
\end{equation}

\noindent 
where $e_{q^\prime}$ is the produced quark (fractional) electric charge, $g$ is the strong charge, 
$a$ and $b$ are the eight-fold color indexes, and
\begin{equation}
  \displaystyle F_1^{\mu \nu} = \Gamma^\mu_{qqZ} \, { \hat p + \hat p_1 + m_{q^\prime}\over (p + p_1)^2 - m_{q^\prime}^2} \, \gamma^\nu + \gamma^\nu { - \hat p - \hat p_2 + m_{q^\prime}\over ( - p - p_2)^2 - m_{q^\prime}^2} \, \Gamma^\mu_{qqZ},
\end{equation}
\begin{equation}
  \displaystyle F_2^{\mu \lambda} = \Gamma^{(+) \, \lambda}_{q^*qg}(k_2,p_1 + p_2) \, { \hat k_1 - \hat p\over (k_1 - p)^2} \, \Gamma^{(-) \, \mu}_{q^*qZ}(k_1,p) \, + \atop { 
    \displaystyle + \, \Gamma^{(+) \, \mu}_{q^*qZ}(k_2,p) { - \hat k_2 - \hat p\over ( - k_2 - p)^2} \, \Gamma^{(-) \, \lambda}_{q^*qg}(k_1,p_1 + p_2) + \Delta^{\mu \lambda}(k_1,-k_2,p,p_1+p_2). }
\end{equation}

\noindent 
The on-shell quark coupling to the $Z$ boson is taken in a standard form:
\begin{equation}
  \Gamma^\mu_{qqZ} = \gamma^\mu (C_V^q - C_A^q \gamma^5).
\end{equation}

\noindent 
The effective vertices can be written as\cite{22,23}:
\begin{equation}
  \Gamma^\mu_{q^*\bar q^* g} = \gamma^\mu - \hat k_1 {l_1^\mu \over (l_1 \cdot k_2)} - \hat k_2 {l_2^\mu \over (l_2 \cdot k_1)}.
\end{equation}
\begin{equation}
  \Gamma^{(+) \, \mu}_{q^*qg}(k,q) = \gamma^\mu - \hat k {l_1^\mu \over (l_1 \cdot q)},
\end{equation}
\begin{equation}
  \Gamma^{(-) \, \mu}_{q^*qg}(k,q) = \gamma^\mu - \hat k {l_2^\mu \over (l_2 \cdot q)}.
\end{equation}

\noindent
The corresponding couplings of the off-shell quark or antiquark to usual on-shell 
quark and $Z$ boson are constructed as it was done earlier\cite{26}:
\begin{equation}
  \Gamma^{(\pm) \, \mu}_{q^*qZ}(k,q) = \Gamma^{(\pm) \, \mu}_{q^*qg}(k,q)(C_V^q - C_A^q \gamma^5).
\end{equation}

\noindent
The induced term $\Delta^{\mu \nu}(k_1,k_2,q_1,q_2)$ has the form\cite{27}:
\begin{equation}
  \Delta^{\mu \nu}(k_1,k_2,q_1,q_2) = \hat k_1 {l_1^\mu l_1^\nu \over (q_1 \cdot l_1) (q_2 \cdot l_1)} + \hat k_2 {l_2^\mu l_2^\nu \over (q_1 \cdot l_2) (q_2 \cdot l_2)}.
\end{equation}

\noindent
The summation on the produced $Z$ boson polarizations is carried out with the usual covariant
formula:
\begin{equation}
  \sum \epsilon^{\mu}(p)\,\epsilon^{*\,\nu}(p) = - g^{\mu \nu} + { p^\mu p^\nu\over m_Z^2}.
\end{equation}

\noindent
In according to the $k_T$-factorization
prescription\cite{17,18}, the summation over the polarizations of 
incoming off-shell gluons is carried with
$\sum \epsilon^\mu \epsilon^{* \nu} = {\mathbf k}_T^\mu {\mathbf k}_T^\nu/{\mathbf k}_T^2$.
In the collinear limit, when $|{\mathbf k}_T| \to 0$, 
this expression converges to the ordinary one 
after averaging on the azimuthal angle.
In according to the using of the effective vertices, 
the spin density matrix for off-shell spinors in initial state is taken in the usual
form $\sum u(x_i l_i) \bar u(x_i l_i) = x_i \hat l_i + m$ (where $i = 1$ or $2$ and we omited the spinor indices).
Further calculations are straightforward and 
in other respects follow the standard QCD Feynman rules. 
The evaluation of traces was performed using the 
algebraic manipulation system \textsc{form}\cite{28}. 
We do not list here the obtained lengthy expressions because of lack of space.
The off-shell amplitude of gluon-gluon fusion subprocess~(3)
was derived in our previous paper\cite{29} (see also\cite{30}).

The cross section of any process
in the $k_T$-factorization approach is
calculated as a convolution of the off-shell partonic cross section 
and the unintegrated, or transverse momentum dependent (TMD), parton 
densities in a proton.
The cross sections of subprocesses~(1) and~(2) read:
\begin{equation}
  \displaystyle \sigma = \sum_{q} \int {1\over 256\pi^3 (x_1 x_2 s)^2} \, |\bar {\cal M}_{1,\,2}|^2\, \times \atop 
  \displaystyle \times \, f_q(x_1,{\mathbf k}_{1T}^2,\mu^2) f_q(x_2,{\mathbf k}_{2T}^2,\mu^2) 
  d{\mathbf k}_{1T}^2 d{\mathbf k}_{2T}^2 d{\mathbf p}_{1T}^2 {\mathbf p}_{2T}^2 dy dy_1 dy_2 \,
  {d\phi_1\over 2\pi} {d\phi_2\over 2\pi} {d\psi_1\over 2\pi} {d\psi_2\over 2\pi},
\end{equation}

\noindent
where $f_q(x_i,\mathbf k_{iT}^2,\mu^2)$ is the TMD quark density in a proton,
$y$ is the rapidity of produced $Z$ boson, $s$ is the total energy,
${\mathbf p}_{1T}$, ${\mathbf p}_{2T}$, $y_1$, $y_2$, $\psi_1$ and $\psi_2$ are the 
transverse momenta, rapidities and azimuthal angles of final state quarks, respectively.
The incoming quarks have azimuthal angles $\phi_1$ and $\phi_2$.
The cross sections of subprocess~(3) can be written as:
\begin{equation}
  \displaystyle \sigma = \int {1\over 256\pi^3 (x_1 x_2 s)^2} \, |\bar {\cal M}_{3}|^2\, \times \atop 
  \displaystyle \times \, f_g(x_1,{\mathbf k}_{1T}^2,\mu^2) f_g(x_2,{\mathbf k}_{2T}^2,\mu^2) 
  d{\mathbf k}_{1T}^2 d{\mathbf k}_{2T}^2 d{\mathbf p}_{1T}^2 {\mathbf p}_{2T}^2 dy dy_1 dy_2 
  {d\phi_1\over 2\pi} {d\phi_2\over 2\pi} {d\psi_1\over 2\pi} {d\psi_2\over 2\pi},
\end{equation}

\noindent
where $f_g(x_i,\mathbf k_{iT}^2,\mu^2)$ is the TMD gluon density in a proton,
and ${\cal M}_{3}$ is the off-shell amplitude of subprocess~(3).

Concerning the TMD parton densities in a proton,
we concentrate on the approach based on the CCFM evolution equation\cite{31}.
The CCFM parton shower, based on the principle of color coherence, describes only
the emission of gluons, while real quark emissions are left aside.
It implies that the CCFM equation describes only the distinct evolution of
TMD gluon and valence quarks, while the non-diagonal transitions between quarks and gluons 
are absent. Below we use the TMD gluon and valence quark
distributions which were obtained\cite{32,33} from the numerical solutions of the CCFM equation (namely, set A0).
Following to\cite{34}, we calculate the TMD sea quark density with the approximation, where the sea quarks 
occur in the last gluon-to-quark splitting. 
At the next-to-leading logarithmic
accuracy $\alpha_s (\alpha_s \ln x)^n$, the TMD sea quark distribution can be written as follows\cite{34}:
\begin{equation}
  f_q^{\rm (sea)}(x,{\mathbf k}_T^2,\mu^2) = \int \limits_x^1 {dz \over z} \int d{\mathbf q}_T^2
    {1\over {\mathbf \Delta}^2} {\alpha_s \over 2\pi} P_{qg}(z,{\mathbf q}_T^2,{\mathbf \Delta}^2) f_g(x/z,{\mathbf q}_T^2, \bar \mu^2),
\end{equation}

\noindent
where $z$ is the fraction of the gluon light cone momentum carried out by
the quark, and $\mathbf \Delta = {\mathbf k}_T - z{\mathbf q}_T$. 
The sea quark evolution is driven by the off-shell gluon-to-quark
splitting function $P_{qg}(z,{\mathbf q}_T^2,{\mathbf \Delta}^2)$\cite{35}:
\begin{equation}
  P_{qg}(z,{\mathbf q}_T^2,{\mathbf \Delta}^2) = T_R \left({\mathbf \Delta}^2\over {\mathbf \Delta}^2 + z(1-z)\,{\mathbf q}_T^2\right)^2
    \left[(1 - z)^2 + z^2 + 4z^2(1 - z)^2 {{\mathbf q}_T^2\over {\mathbf \Delta}^2} \right],
\end{equation}

\noindent 
where $T_R = 1/2$. The splitting function $P_{qg}(z,{\mathbf q}_T^2,{\mathbf \Delta}^2)$
has been obtained by generalizing to finite transverse momenta, in the high-energy region, 
the two-particle irreducible kernel expansion\cite{36}.
It takes into account the small-$x$ enhanced transverse momentum dependence 
up to all orders in the strong coupling constant, and reduces to the conventional splitting
function at lowest order for $|\mathbf q_T| \to 0$.
The scale $\bar \mu^2$ is defined\cite{37} from the angular ordering condition which is natural
from the point of view of the CCFM evolution: $\bar \mu^2 = {\mathbf \Delta}^2/(1-z)^2 + {\mathbf q}_T^2/(1-z)$.

Other essential parameters were taken as follows: renormalization scale
$\mu_R^2 = m_Z^2 + p_T^2$, factorization scale $\mu_F^2 = \hat s + {\mathbf Q}_T^2$
(with ${\mathbf Q}_T$ being the transverse momentum of initial parton pair),
beauty quark mass $m_b = 4.75$~GeV, $m_Z = 91.1876$~GeV, $m_H = 125$~GeV, $\Gamma_Z = 2.4952$~GeV, $\Gamma_H = 4.3$~MeV,
$\sin^2 \theta_W = 0.23122$ and we use the LO formula for the strong coupling constant
$\alpha_s(\mu^2)$ with $n_f= 4$ active quark flavors at
$\Lambda_{\rm QCD} = 200$~MeV, so that $\alpha_s (m_Z^2) = 0.1232$.
To take into account the non-logarithmic loop corrections to the 
production cross sections, we apply the effective $K$-factor,
as it was done in\cite{38,39}:
\begin{equation}
  K = \exp \left[ C_F {\alpha_s(\mu^2)\over 2\pi} \pi^2 \right],
\end{equation}

\noindent
where color factor $C_F = 4/3$. A particular scale choice
$\mu^2 = p_T^{4/3} m_Z^{2/3}$ was proposed\cite{38,39}
to eliminate sub-leading logarithmic terms. Note we choose this scale
to evaluate the strong coupling constant in~(22) only.
Everywhere the multidimensional integration have been performed
by the means of Monte Carlo technique, using the routine \textsc{vegas}\cite{40}.
The corresponding C++ code is available from the authors on request\footnote{lipatov@theory.sinp.msu.ru}.

We now are in a position to present our numerical predictions.
The differential cross sections of associated $Zb\bar b$ production in 
$pp$ collisions as a function of $M$, the
invariant mass of final beauty quarks,
and $Z$ boson transverse momentum 
at $\sqrt s = 8$ and $14$~TeV are shown in Figs.~2 and~3. 
The solid, dashed and dash-dotted histograms correspond to the contributions from the 
subprocesses~(1), (2) and (3), respectively.
There is 
no any cuts applied. One can see that the associated Higgs (decaying into the $b\bar b$ pair) 
and $Z$ boson production 
cross section lies below the QCD backgrounds
by several orders of magnitude in a whole $p_T$ range,
but peaks near Higgs mass.
To increase the relative contribution from Higgs signal,
we repeated the calculations in the restricted region of $M$,
namely $120 < M < 130$~GeV (see Fig.~4).
We found that here the associated Higgs and $Z$ boson production
gives a sizeble contribution to the $Zb\bar b$ cross section
at high $Z$ boson transverse momenta. 
So, at $\sqrt s = 8$~TeV it 
practically coincides with the leading contribution from the 
gluon-gluon fusion subprocess at $p_T > 200$~GeV.
At $\sqrt s = 14$~TeV, it lies below the latter. However, these
contributions are almost comparable at $p_T > 300$~GeV.
With the expected LHC luminosity of about $40$~fb$^{-1}$,
our estimation gives $400$ --- $500$ events (with 
beauty quarks originating from the Higgs boson decays) for both energies, $8$ and $14$~TeV.
Therefore, the possibility for the experimental detection 
of Higgs signal appears in the kinematical region defined above.

A special opportunity to detect the decays of scalar Higgs bosons 
can be provided by the investigations of different angular correlations 
between the final state particles. 
As an example, we calculated the distributions on the angle $\theta$ between the
produced $Z$ boson and $b$-quark in the Collins-Soper frame (where
$z$ axis is defined with respect to the bisector of colliding 
protons in the $b\bar b$ rest frame), and
on the azimuthal angle difference $\Delta \phi$
between the final beauty quarks in the $pp$ center-of-mass frame.
The results of our calculations
performed near Higgs boson peak (with $120 < M < 130$~GeV)
are shown in Figs.~5 and~6, where an additional cut $p_T > 200$($300$)~GeV is applied
at $\sqrt s = 8$($14$)~TeV.
As it was expected, the isotropic decay of scalar Higgs particle $H \to b\bar b$
greatly differs from the angular distributions
predicted by the off-shell amplitudes of subprocesses~(2) and~(3).
Moreover, the beauty quarks, originating from the Higgs boson decay, populate
mostly at low $\Delta \phi$ (see Fig.~6), whereas the leading QCD 
background, as given by the gluon-gluon fusion subprocess~(3), has more flat $\Delta \phi$ distribution.
So, the different angular correlations 
between the final state particles
in the associated $Zb\bar b$ production
are very sensitive to the source of $b\bar b$ pairs,
and therefore future experimental investigations of 
such observables at the LHC with increased luminosity can  
give a clear information about Higgs signal.

Finally, we study the size of theoretical uncertainties of our calculations
connected with the hard scale. As usual, in order to estimate these uncertainties
we vary the scales by a factor of $2$ around their default values.
Also, we use the CCFM set A0$+$ and A0$-$ instead of the default TMD gluon density 
A0. These two PDF sets represent a variation of the hard
scale involved in~(18) and~(19). The A0$+$ stands for a variation of $2\mu$, while
set A0$-$ reflects $\mu/2$. We observe a deviation of about $50$\%
with both A0$+$ and A0$-$ sets (see Fig.~7)
for the QCD 
background (as given by the sum of gluon-gluon 
fusion and quark-antiquark annihilation subprocesses considered above).
Despite the relatively large band of uncertainties, the latter does not change our
conclusions. Additionally, to investigate the role of 
higher-order QCD corrections, in Fig.~7 we presented the results for the 
QCD background obtained in the framework of collinear QCD factorization at LO.
We find that in the kinematical
region where the possible Higgs signal could be observed 
these corrections are important.  

To conclude, in the present note we applied the 
$k_T$-factorization approach of QCD to 
study the possibility to 
detect the scalar Higgs boson decay $H\to b\bar b$  
in the associated $Z$ and $b\bar b$ production at the LHC.
Our consideration was based on the off-shell production 
amplitudes of $q^* \bar q^* \to Z H \to Z q^\prime \bar q^\prime$,
$q^* \bar q^* \to Z q^\prime \bar q^\prime$ and
$g^* g^* \to Z q^\prime \bar q^\prime$ 
partonic subprocesses supplemented with the CCFM dynamics of parton densities in a proton.
The main part of higher-order QCD corrections
(corresponding to the $\log 1/x$ enhanced 
terms in perturbative series) is effectively taken into account
in our consideration.
We demonstrated that 
the $H \to b\bar b$ signal can be observed at large transverse 
momenta near Higgs boson peak despite the overwhelming QCD background, and pointed out an important 
role of angular correlations between the produced $Z$ 
boson and $b$-quarks.
The gauge invariant off-shell amplitudes of
$q^* \bar q^* \to Z H \to Z q^\prime \bar q^\prime$ and
$q^* \bar q^* \to Z q^\prime \bar q^\prime$ 
partonic subprocesses, calculated for the first time, 
can be implemented in a different Monte Carlo
event generators, like as, for example, \textsc{cascade}\cite{41}.

{\sl Acknowledgements.}
The authors are grateful to H.~Jung
for very useful discussions which give us the idea of present study.
We thank also S.~Baranov for helpful discussion concerning the role
of angular distributions and M.~Malyshev for additional
check of the off-shell amplitude of $q^* \bar q^* \to ZH \to Z q^\prime \bar q^\prime$
subprocess.
This research was supported by the FASI of Russian Federation
(grant NS-3042.2014.2).
We are also grateful to DESY Directorate for the
support in the framework of Moscow---DESY project on Monte-Carlo implementation for
HERA---LHC.

\newpage

\begin{figure}
\begin{center}
\epsfig{figure=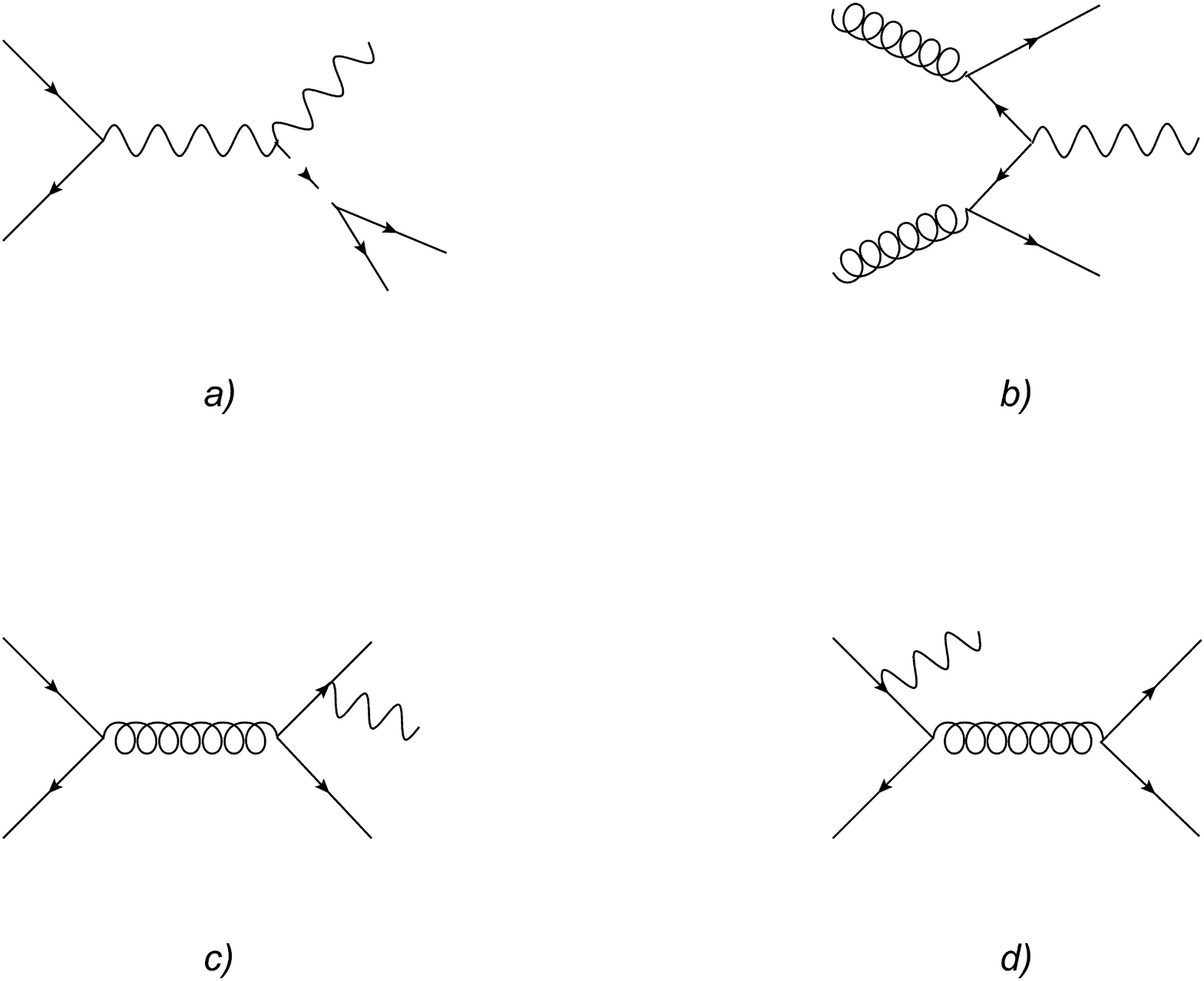, width = 11cm}
\caption{Examples of Feynman diagrams corresponding to 
$q^*\bar q^* \to Z H \to Z b \bar b$ {\sl (a)}, 
$g^*g^* \to Z b \bar b$ {\sl (b)} and $q^* \bar q^* \to Z b \bar b$ {\sl (c, d)} subprocesses.
Full set of diagrams can be obtained by permutations of quark, gluon, Higgs and $Z$ boson lines.}
\label{fig1}
\end{center}
\end{figure}

\begin{figure}
\begin{center}
\epsfig{figure=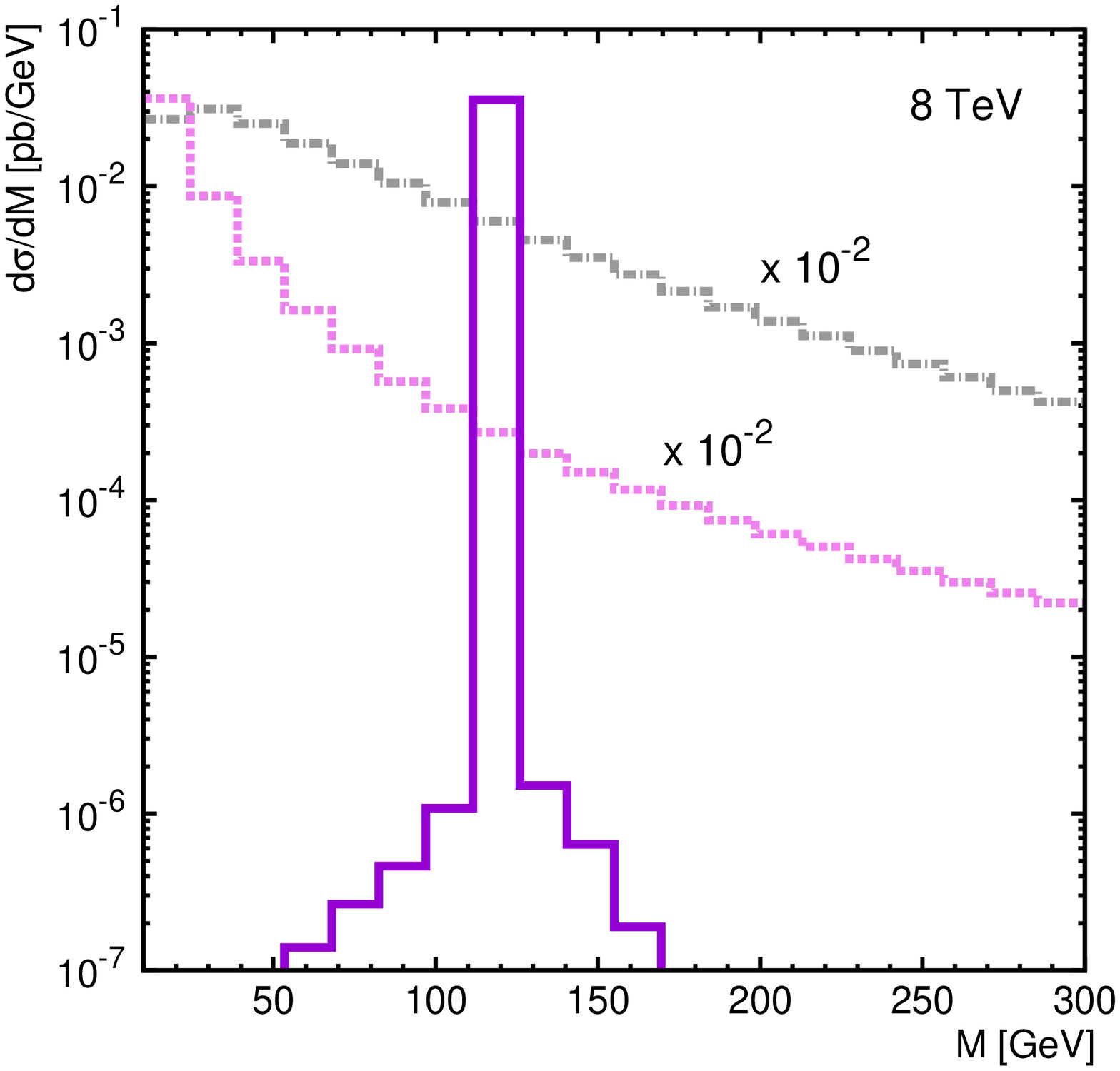, width = 5.5cm, angle = 270}
\epsfig{figure=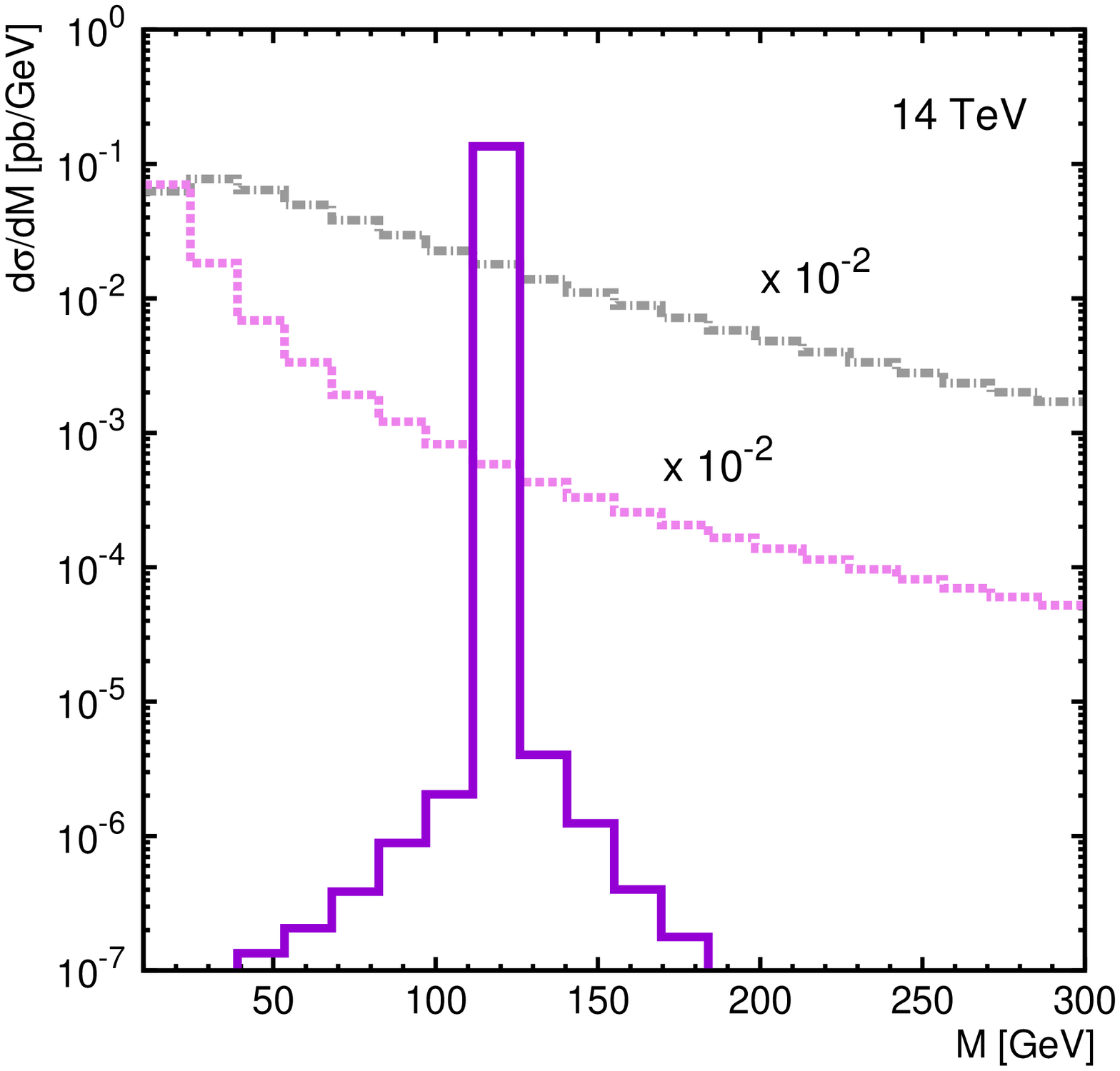, width = 5.5cm, angle = 270}
\caption{The associated $Z + b \bar b$ cross sections 
in $pp$ collisions calculated as a function of invariant mass of $b\bar b$ quarks 
at $\sqrt s = 8$~TeV (left panel) and $\sqrt s = 14$~TeV (right panel).
The solid, dashed and dash-dotted histograms correspond to the contributions from the 
$q^*\bar q^* \to Z H \to Z b \bar b$,
$q^* \bar q^* \to Z b \bar b$ and $g^*g^* \to Z b \bar b$ subprocesses, respectively.
No cuts is applied.}
\label{fig2}
\end{center}
\end{figure}

\begin{figure}
\begin{center}
\epsfig{figure=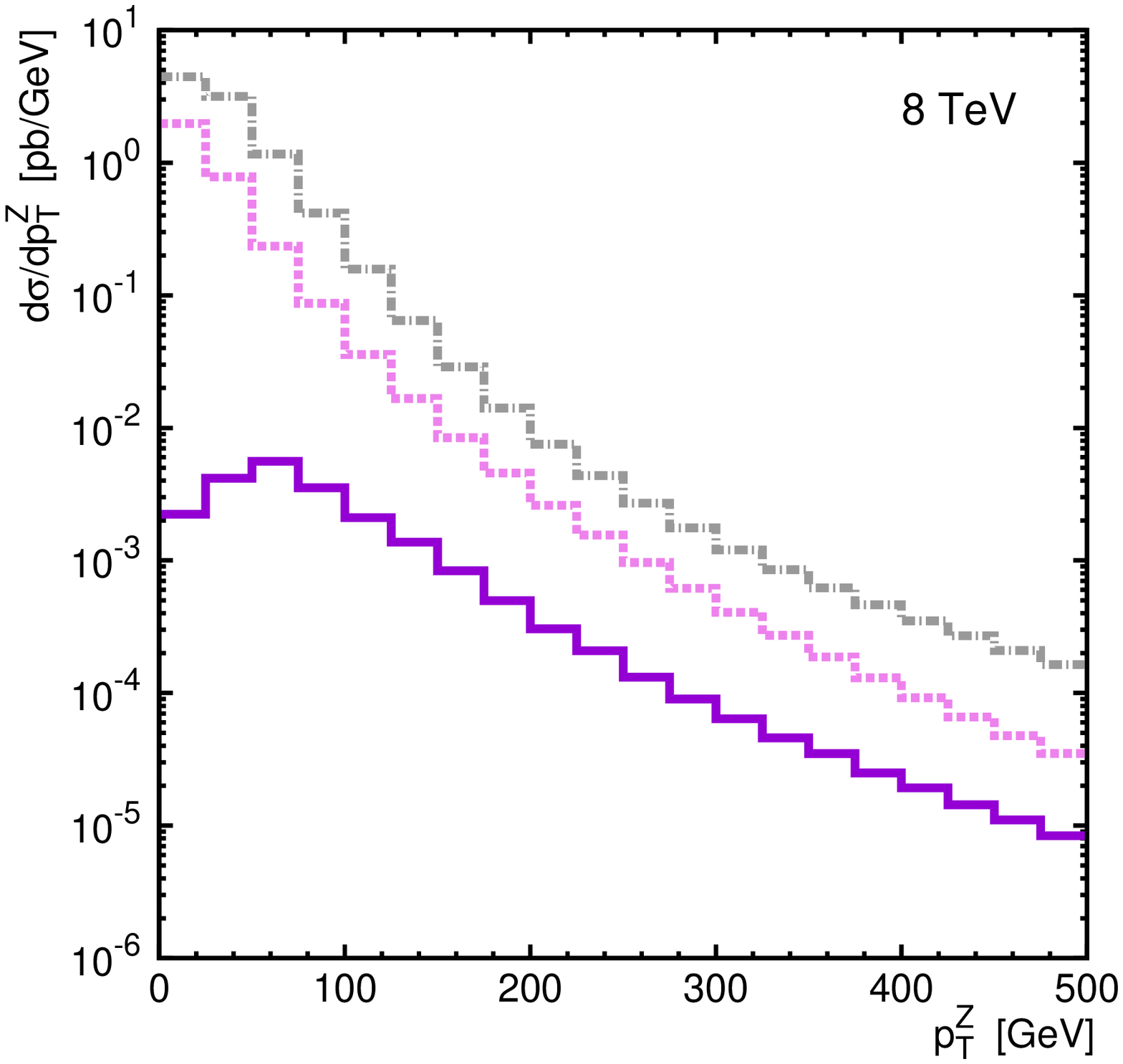, width = 5.5cm, angle = 270}
\epsfig{figure=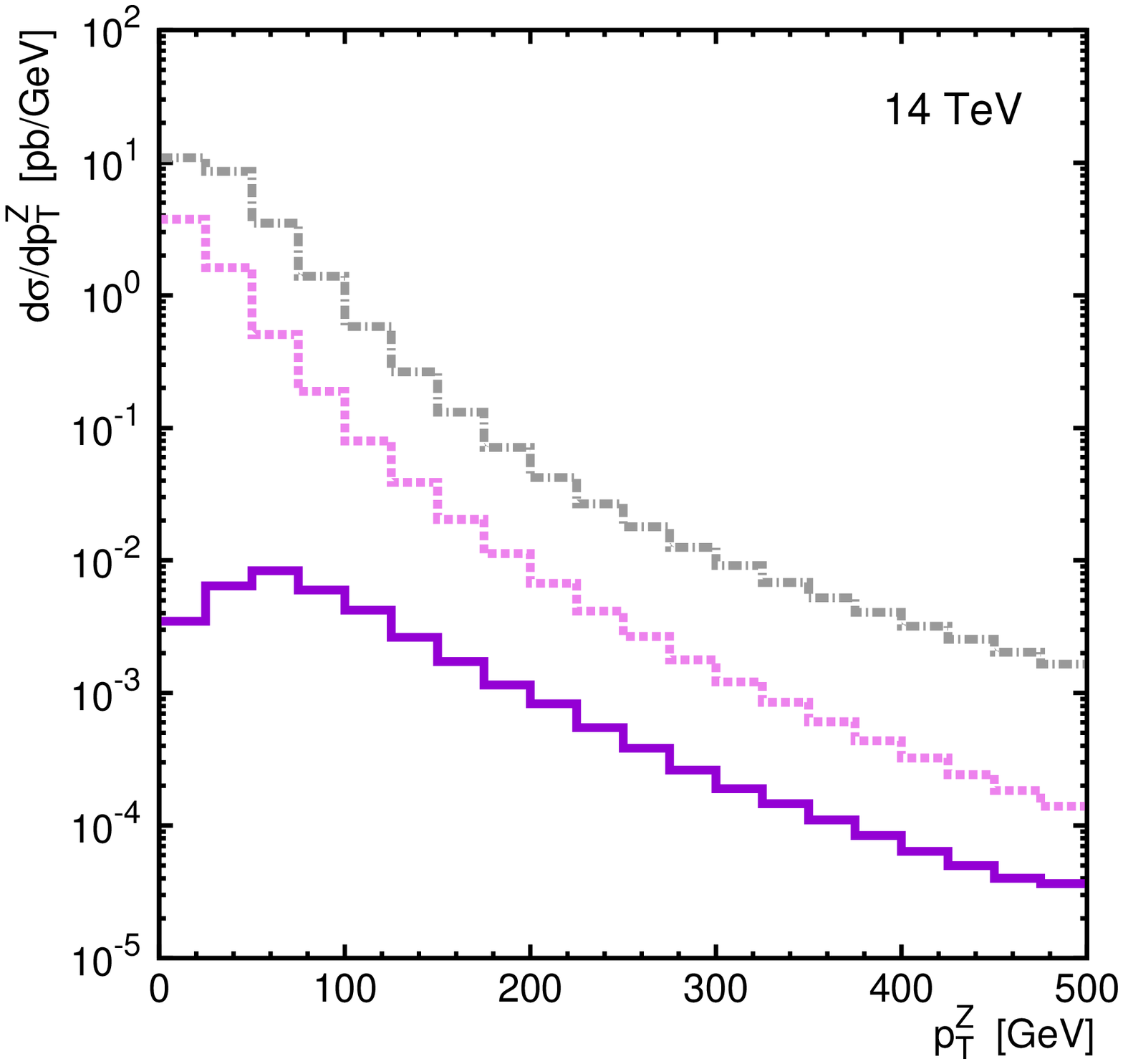, width = 5.5cm, angle = 270}
\caption{The associated $Z + b \bar b$ cross sections 
in $pp$ collisions calculated as a function of $Z$ boson transverse momentum at $\sqrt s = 8$~TeV
(left panel) and $\sqrt s = 14$~TeV (right panel).
Notation of all histograms is the same
as in Fig.~2. No cuts is applied.}
\label{fig3}
\end{center}
\end{figure}

\begin{figure}
\begin{center}
\epsfig{figure=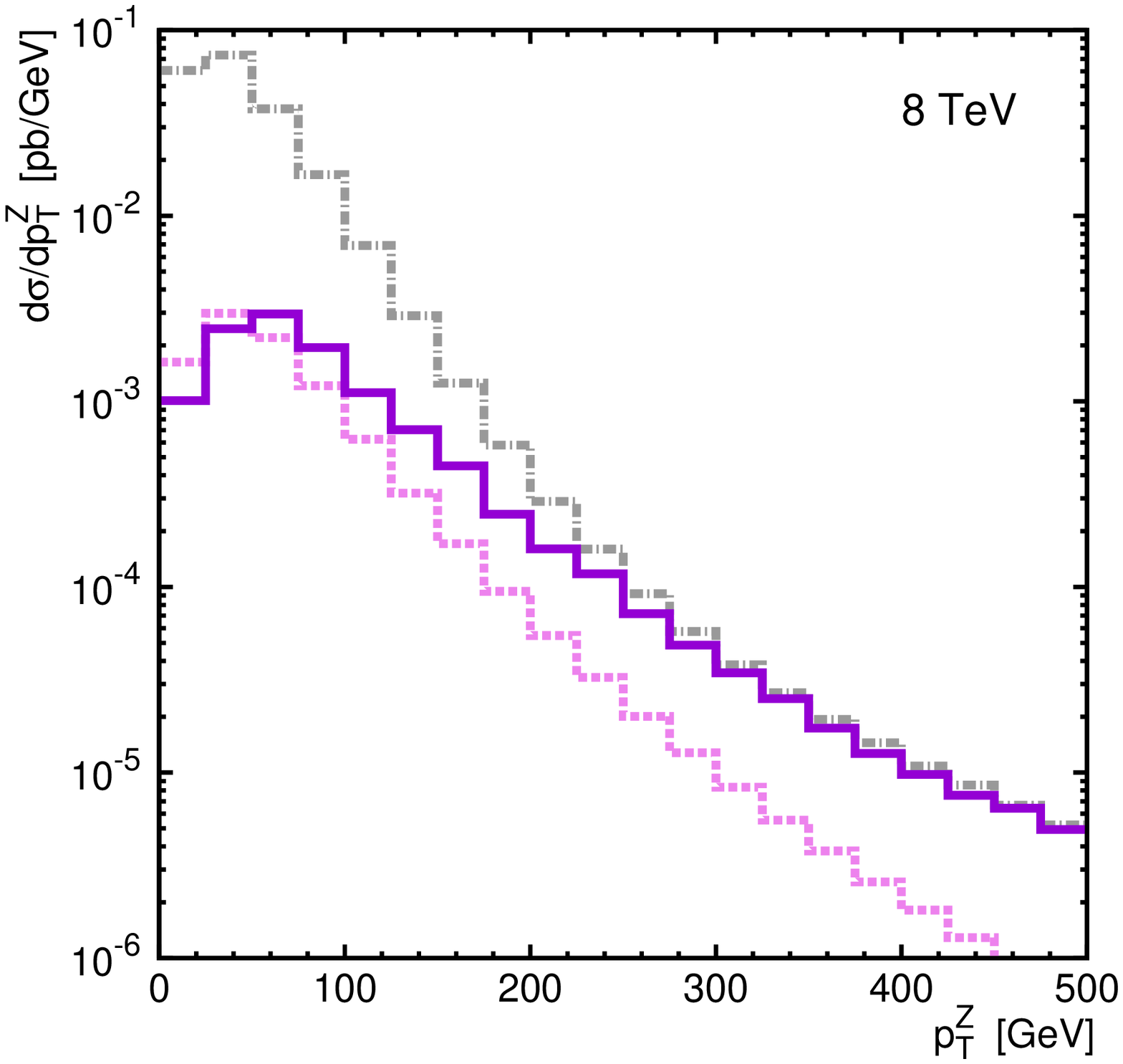, width = 5.5cm, angle = 270}
\epsfig{figure=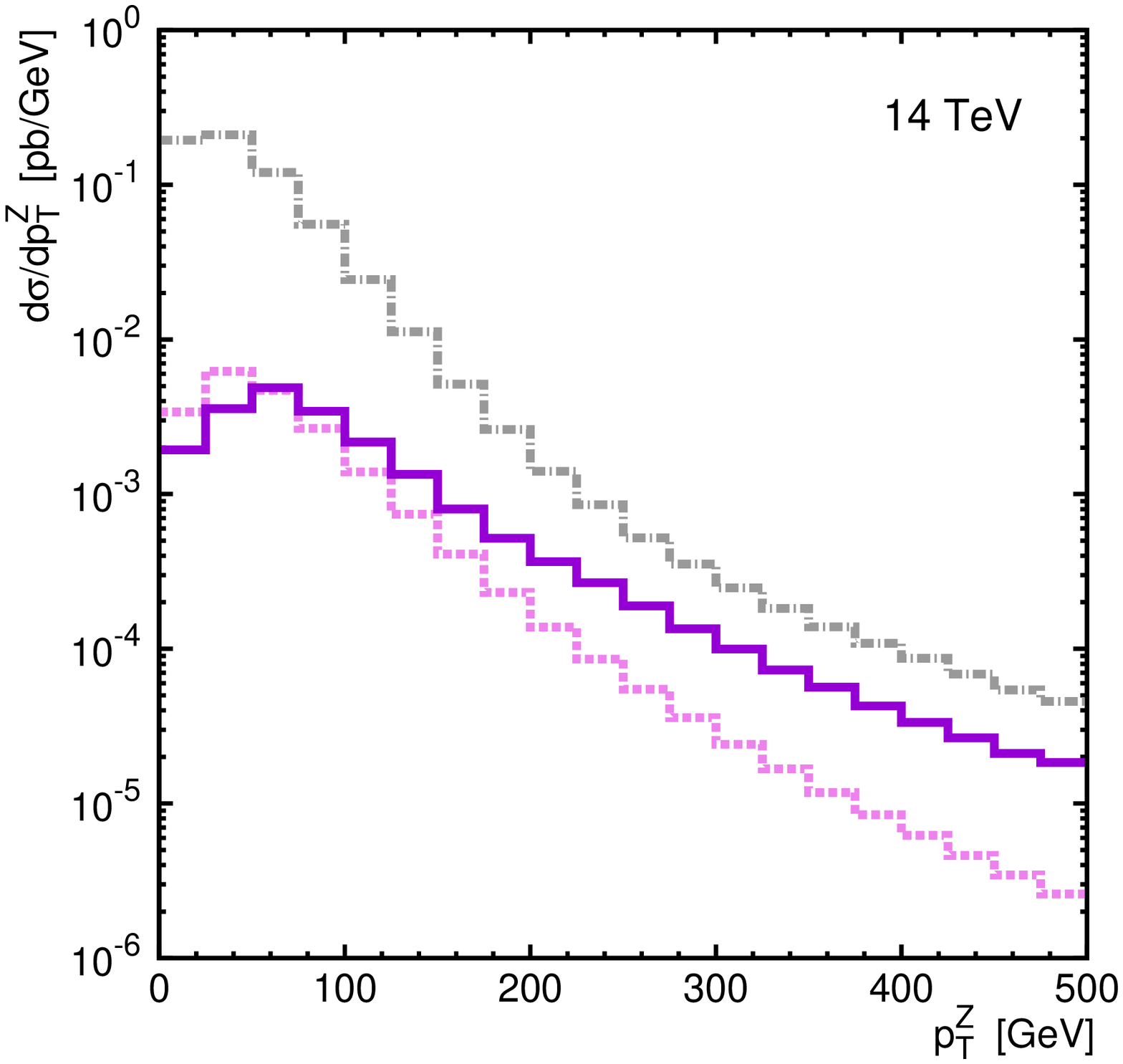, width = 5.5cm, angle = 270}
\caption{The associated $Z + b \bar b$ cross sections 
in $pp$ collisions calculated as a function of $Z$ boson transverse momentum at $\sqrt s = 8$~TeV
(left panel) and $\sqrt s = 14$~TeV (right panel) at $120 < M < 130$~GeV.
Notation of all histograms is the same
as in Fig.~2.}
\label{fig4}
\end{center}
\end{figure}

\begin{figure}
\begin{center}
\epsfig{figure=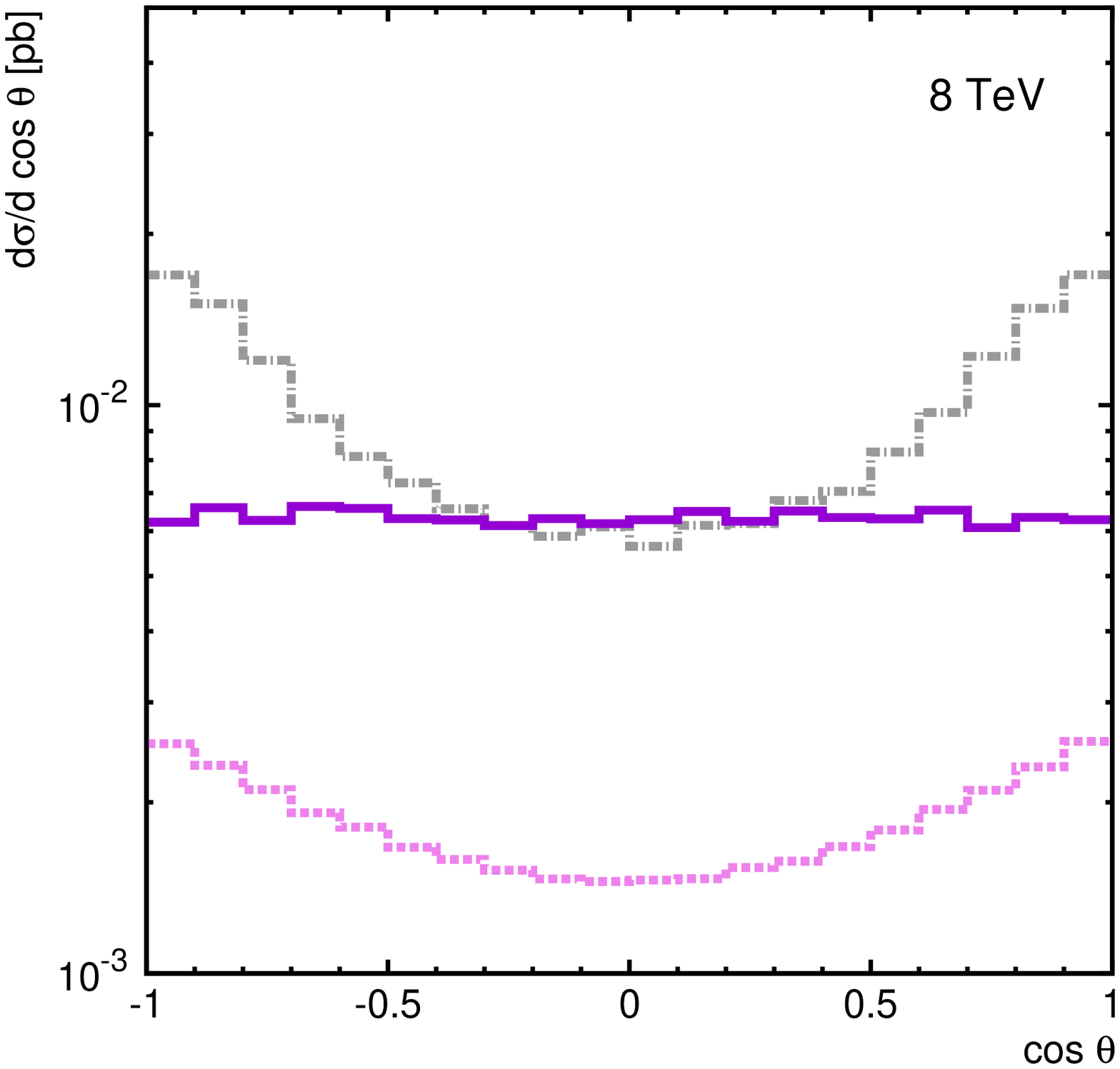, width = 5.5cm, angle = 270}
\epsfig{figure=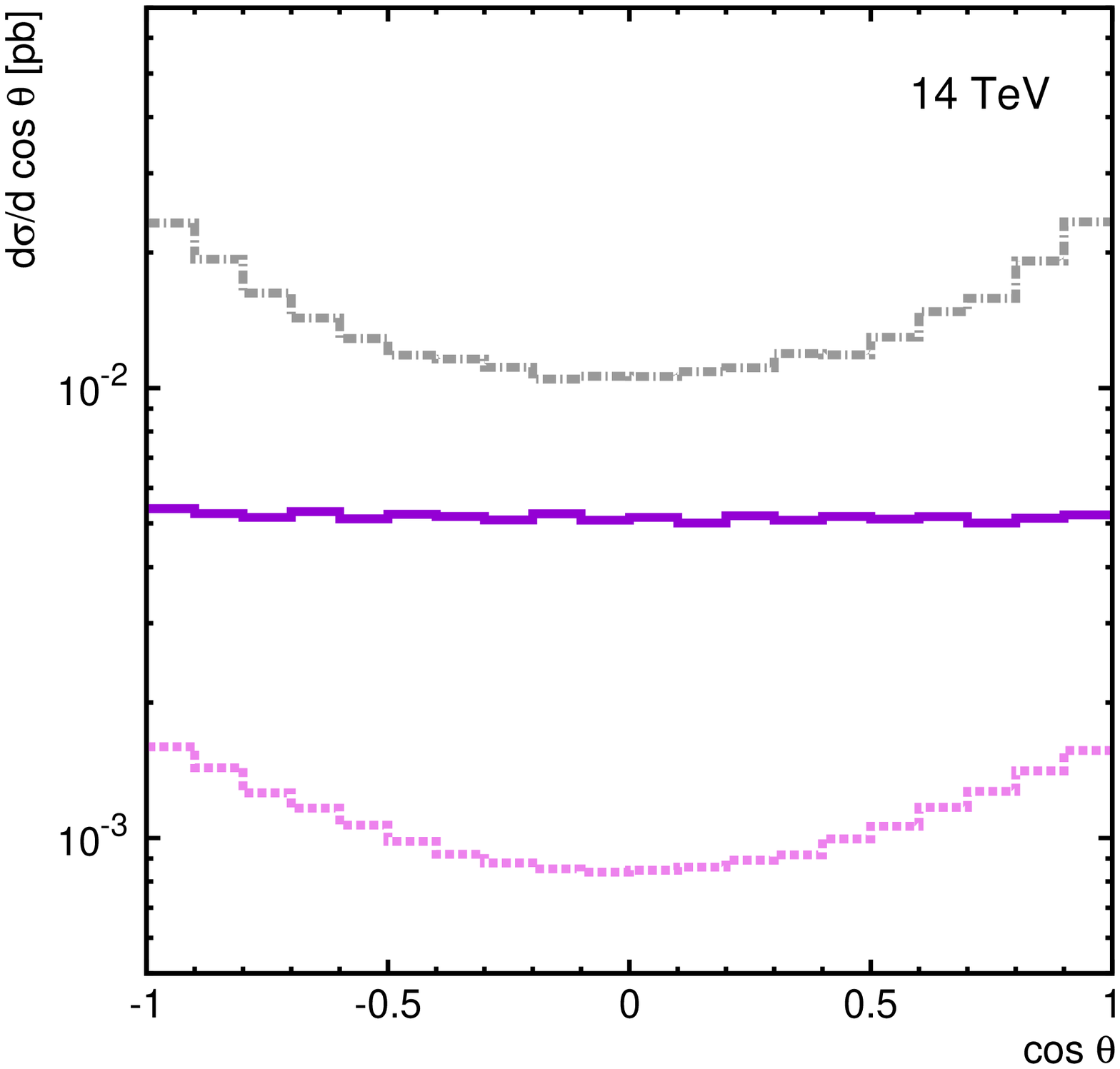, width = 5.5cm, angle = 270}
\caption{The associated $Z + b \bar b$ cross sections 
in $pp$ collisions calculated as a function of angle $\theta$ between the
produced $Z$ boson and beauty quark in the Collins-Soper frame
at $\sqrt s = 8$~TeV (left panel) and $\sqrt s = 14$~TeV (right panel) at $120 < M < 130$~GeV.
An additional cut $p_T > 200$($300$)~GeV is applied for $\sqrt s = 8$($14$)~TeV.
Notation of all histograms is the same
as in Fig.~2.}
\label{fig5}
\end{center}
\end{figure}

\begin{figure}
\begin{center}
\epsfig{figure=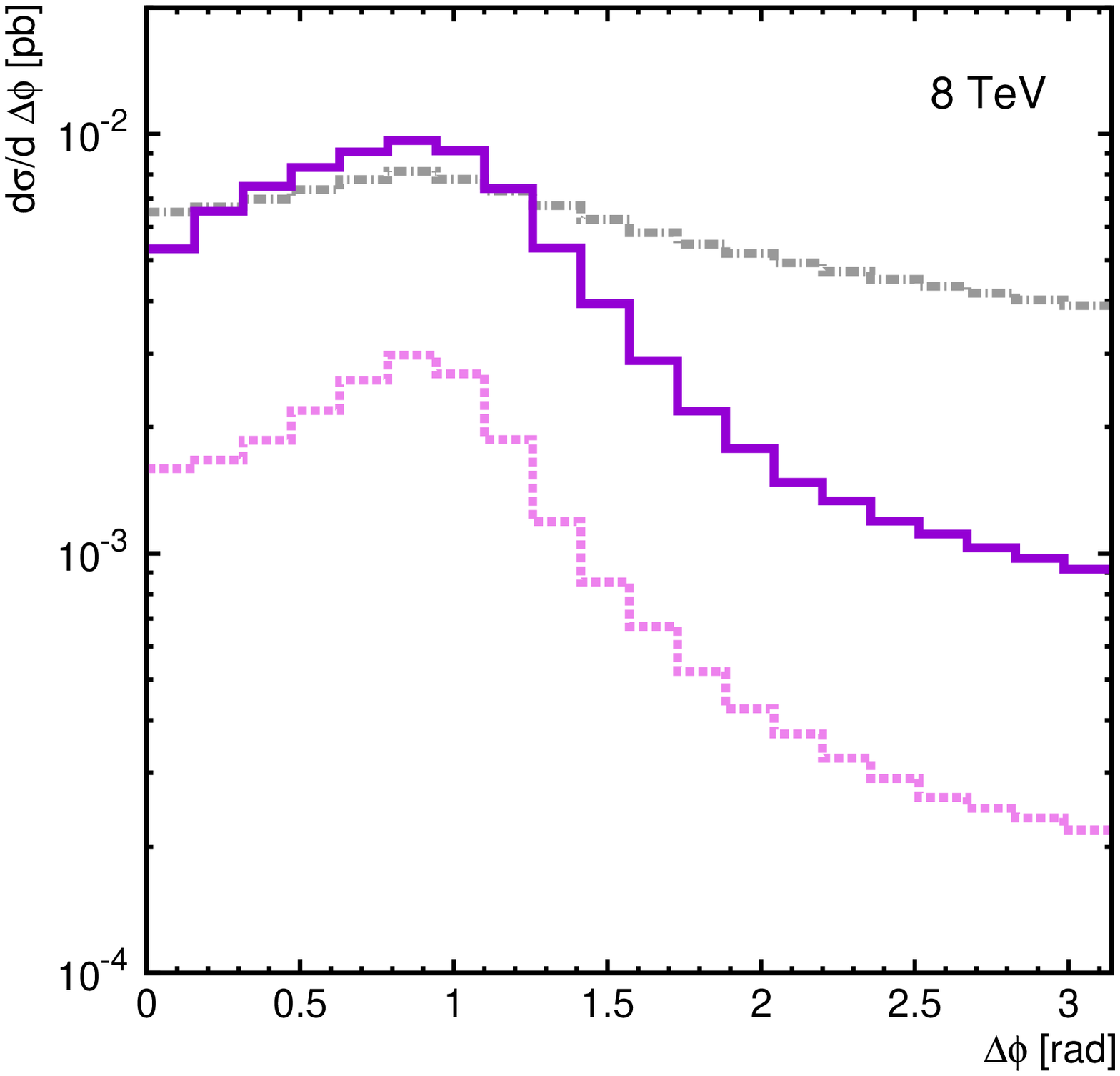, width = 5.5cm, angle = 270}
\epsfig{figure=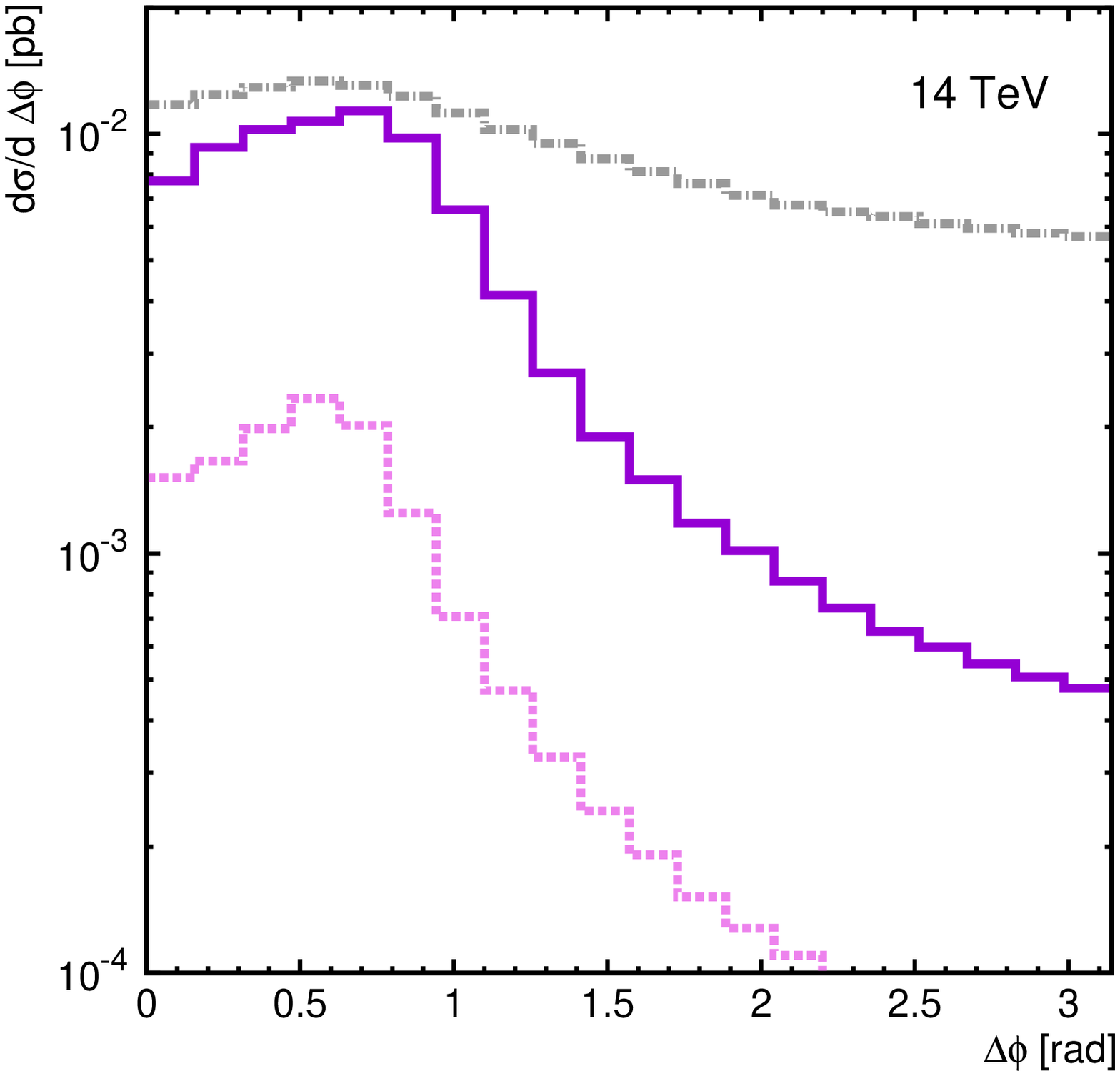, width = 5.5cm, angle = 270}
\caption{The associated $Z + b \bar b$ cross sections 
in $pp$ collisions calculated as a function of azimuthal angle difference $\Delta \phi$ between the
produced beauty quarks in the $pp$ center-of-mass frame
at $\sqrt s = 8$~TeV (left panel) and $\sqrt s = 14$~TeV (right panel) at $120 < M < 130$~GeV.
An additional cut $p_T > 200$($300$)~GeV is applied for $\sqrt s = 8$($14$)~TeV.
Notation of all histograms is the same
as in Fig.~2.}
\label{fig6}
\end{center}
\end{figure}

\begin{figure}
\begin{center}
\epsfig{figure=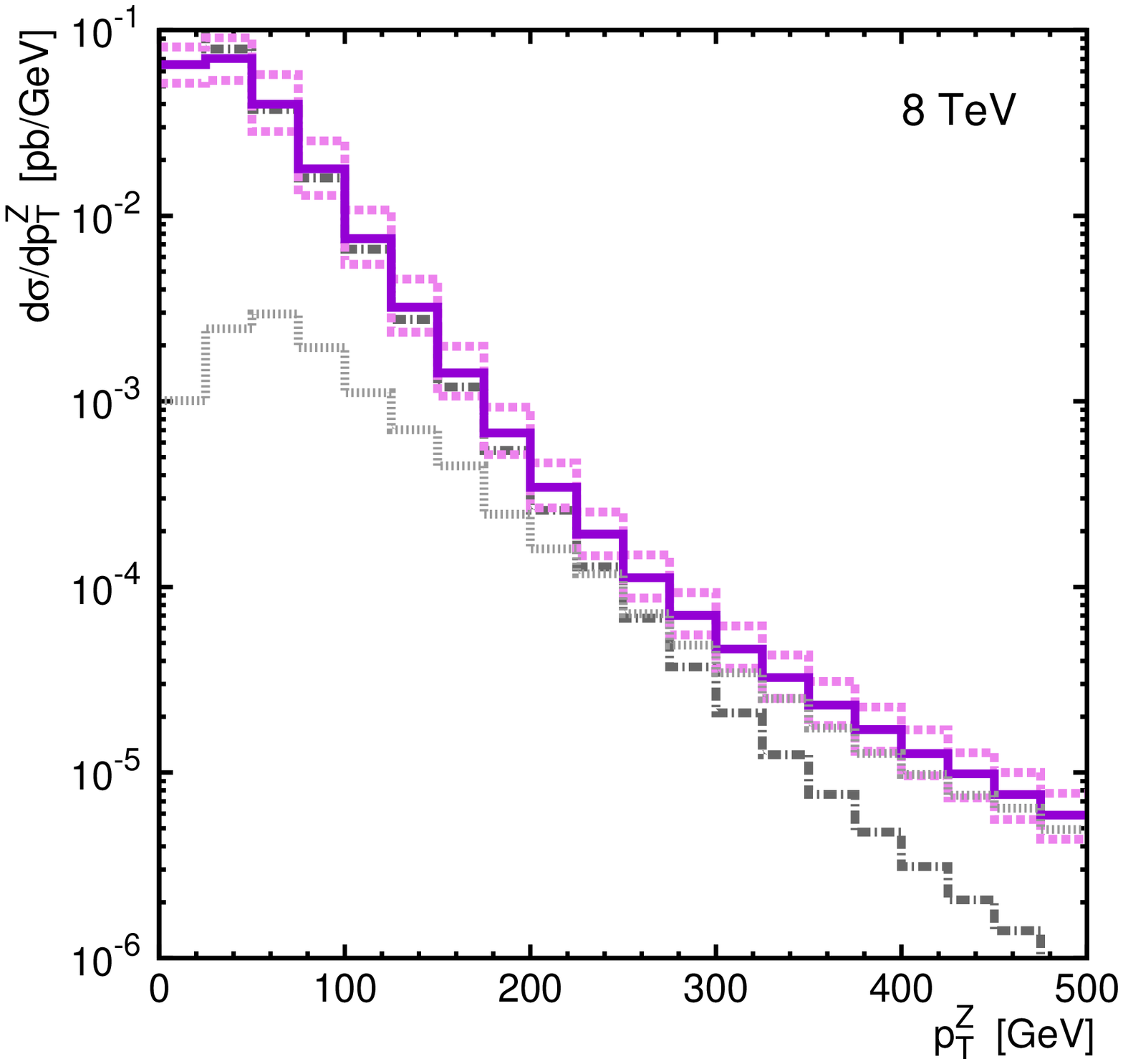, width = 5.5cm, angle = 270}
\epsfig{figure=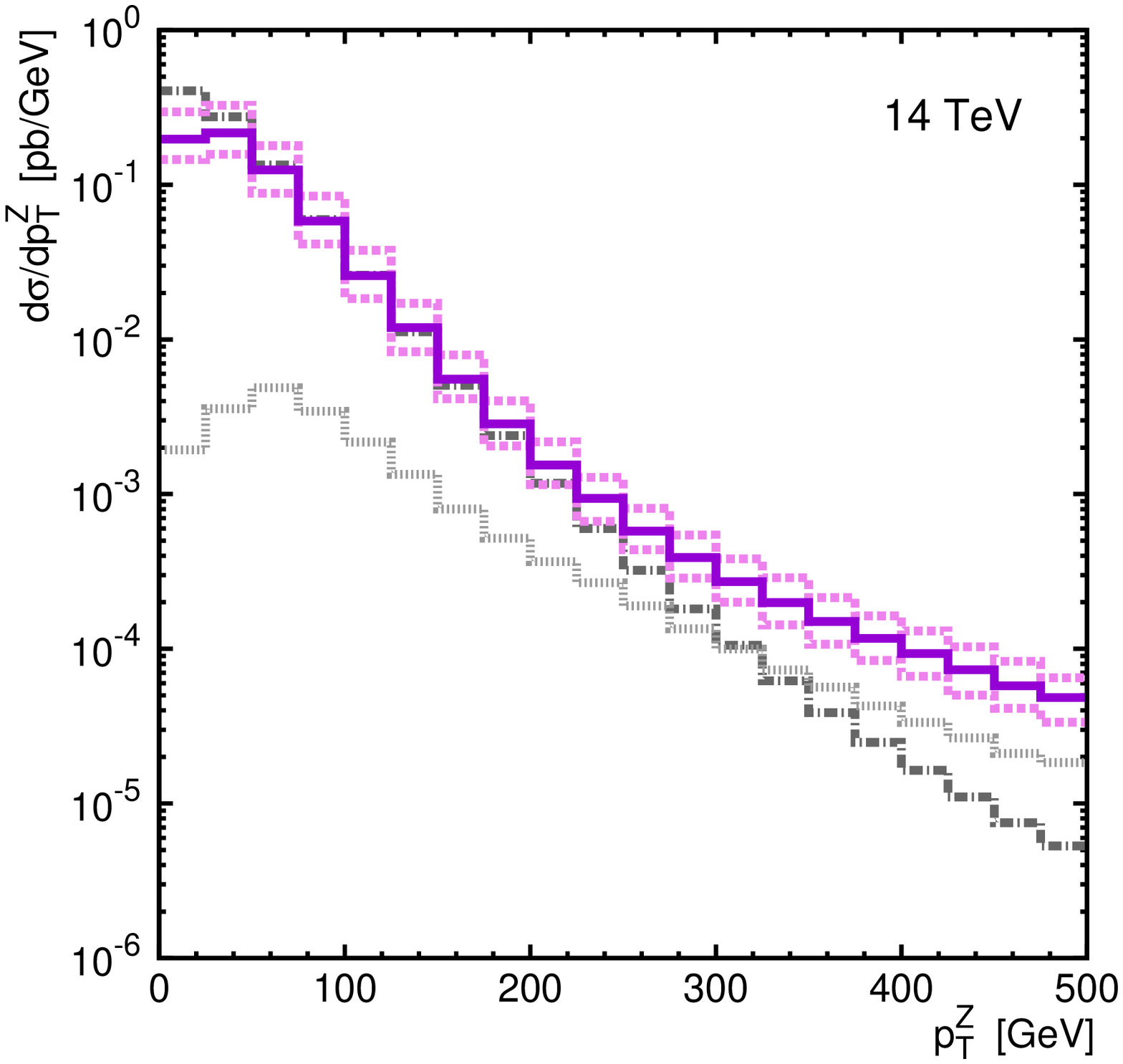, width = 5.5cm, angle = 270}
\epsfig{figure=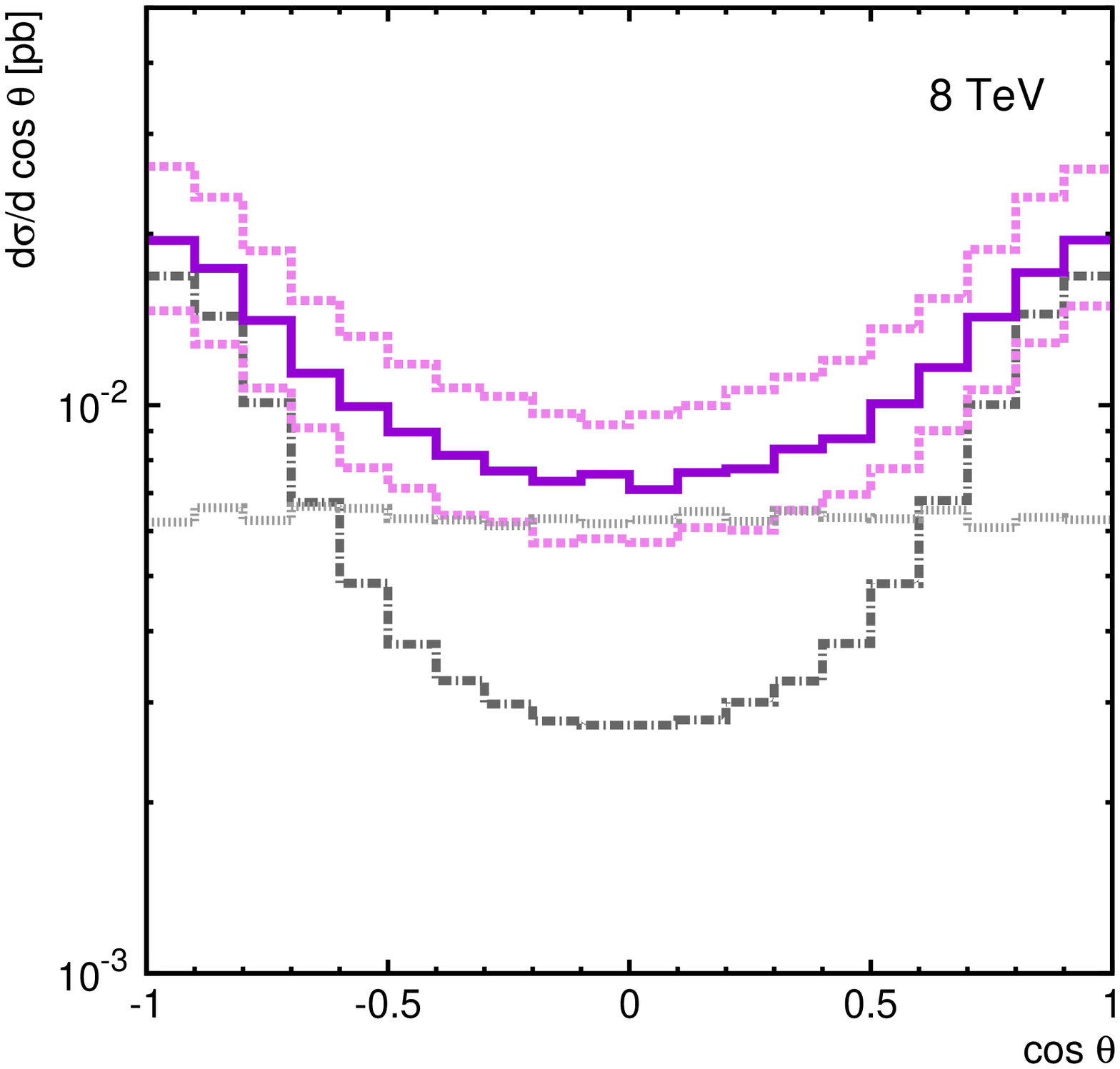, width = 5.5cm, angle = 270}
\epsfig{figure=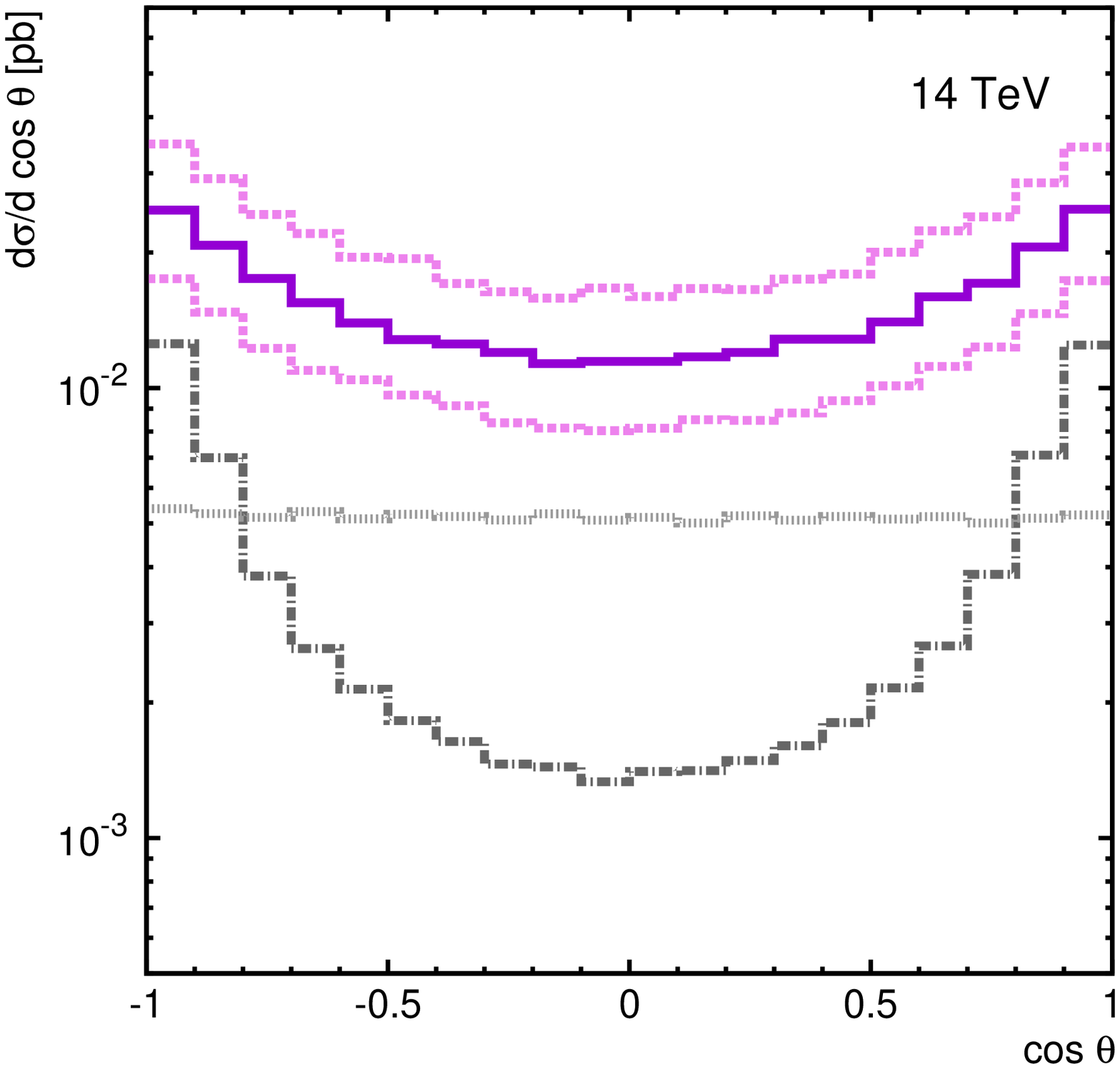, width = 5.5cm, angle = 270}
\epsfig{figure=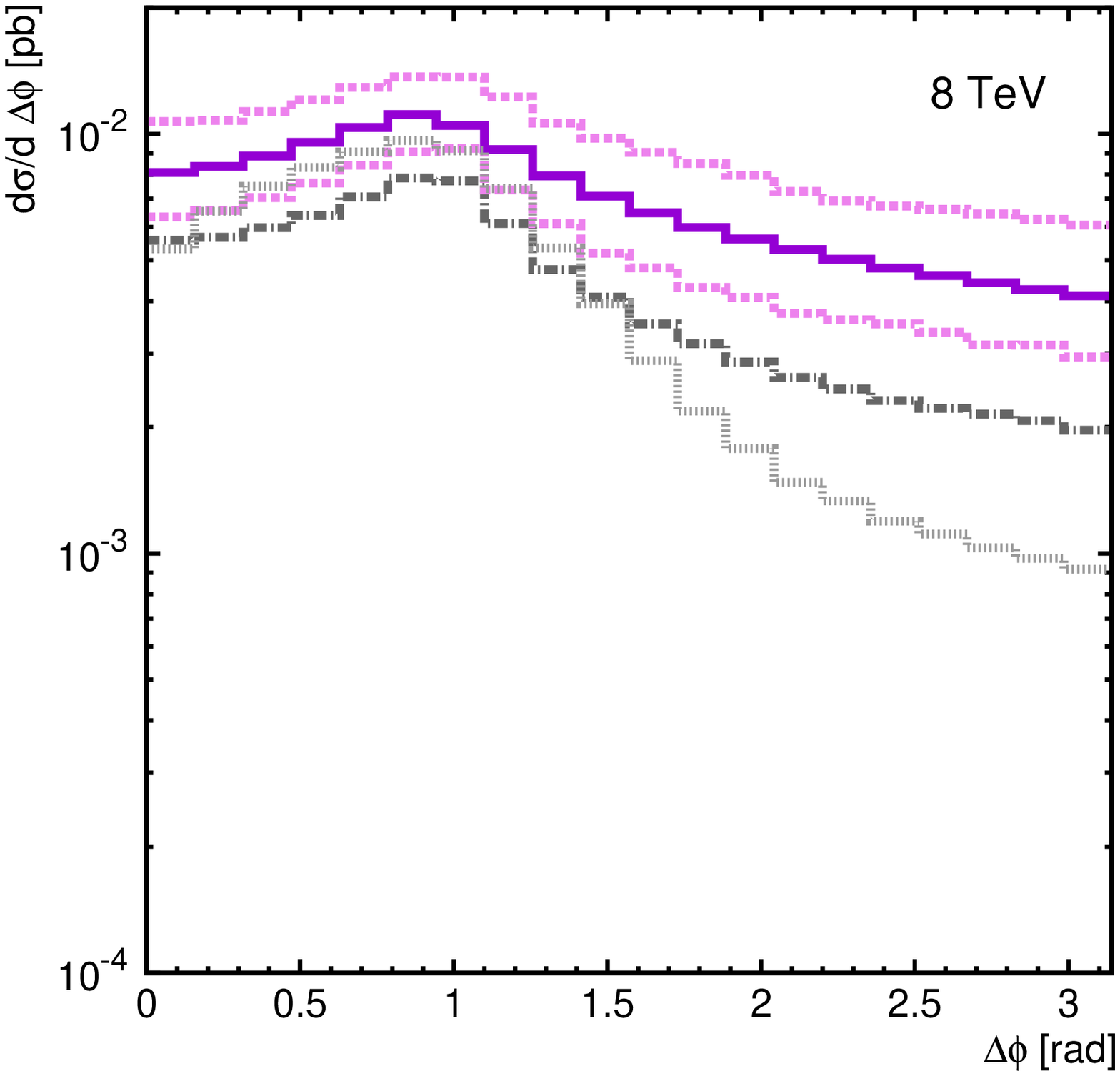, width = 5.5cm, angle = 270}
\epsfig{figure=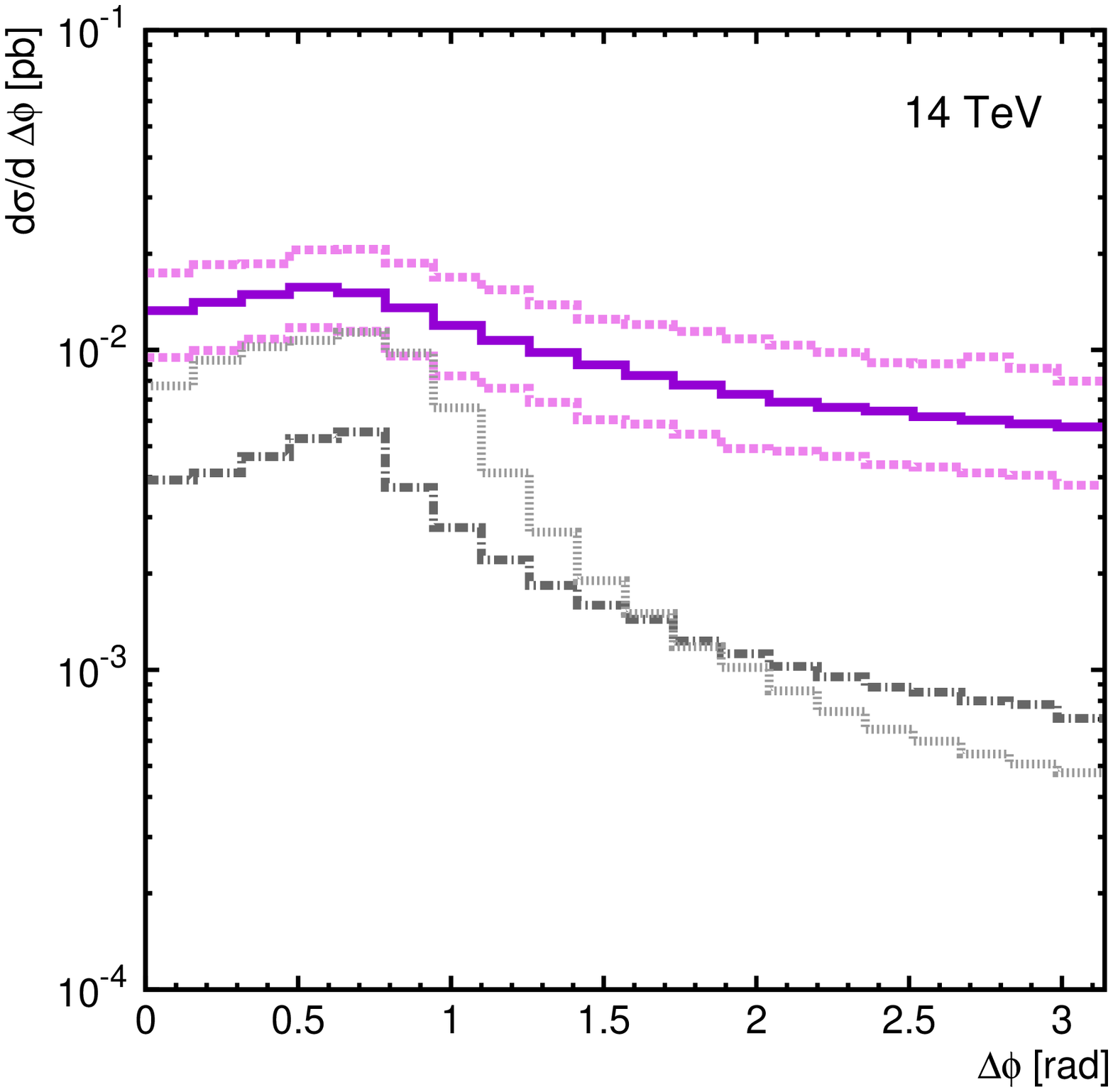, width = 5.5cm, angle = 270}
\caption{The associated $Z + b \bar b$ cross sections 
in $pp$ collisions calculated as a function of $Z$ boson transverse momentum $p_T$, angle $\theta$ and
azimuthal angle difference $\Delta \phi$ at $\sqrt s = 8$~TeV (left panel) and $\sqrt s = 14$~TeV (right panel) 
at $120 < M < 130$~GeV. The solid and dash-dotted histograms correspond to the QCD background
(sum of the gluon-gluon fusion and quark-antiquark annihilation subprocesses)
calculated in the framework of $k_T$-factorization approach and collinear approximation of QCD at LO, respectively.
The upper and lower dashed histograms correspond to the scale variations 
in the $k_T$-factorization predictions, as it is described in the text.
The dotted histograms correspond to the contributions from the $q^*\bar q^* \to Z H \to Z b \bar b$
subprocess. An additional cut $p_T > 200$($300$)~GeV is applied for $\sqrt s = 8$($14$)~TeV 
in the $\theta$ and $\Delta \phi$ distributions.}
\label{fig7}
\end{center}
\end{figure}


\begin{thebibliography}{41}
\bibitem{1} CMS Collaboration, Phys. Lett. B {\bf 716}, 30 (2012).
\bibitem{2} ATLAS Collaboration, Phys. Lett. B {\bf 716}, 1 (2012).
\bibitem{3} ATLAS Collaboration, JHEP {\bf 09}, 112 (2014).
\bibitem{4} M.~Spira, A.~Djouadi, D.~Graudenz, P.~Zerwas, Nucl. Phys. B {\bf 453} 17, (1995).
\bibitem{5} A.~Djouadi, M.~Spira, P.M.~Zerwas, Phys. Lett. B {\bf 264} 440, (1991).
\bibitem{6} S.~Dawson, Nucl. Phys. B {\bf 359} 283, (1991).
\bibitem{7} R.V.~Harlander, W.B.~Kilgore, Phys. Rev. Lett. {\bf 88}, 201801 (2002).
\bibitem{8} C.~Anastasiou, K.~Melnikov, Nucl. Phys. B {\bf 646}, 220 (2002).
\bibitem{9} V.~Ravindran, J.~Smith, W.L.~van~Neerven, Nucl. Phys. B {\bf 665}, 325 (2003).
\bibitem{10} S.~Catani, D.~de~Florian, M.~Grazzini, P.~Nason, JHEP {\bf 0307}, 028 (2003).
\bibitem{11} D.~de~Florian, G.~Ferrera, M.~Grazzini, D.~Tommasini, JHEP {\bf 1111}, 064 (2011).
\bibitem{12} CMS Collaboration, Phys. Rev. D {\bf 89}, 092007 (2014).
\bibitem{13} CMS Collaboration, JHEP {\bf 01}, 096 (2014).
\bibitem{14} LHC Higgs Cross Section Working Group Collaboration, CERN-2011-002.
\bibitem{15} CDF and D0 Collaborations, Phys. Rev. Lett. {\bf 109}, 071804 (2012).
\bibitem{16} CMS Collaboration, Phys. Rev. D {\bf 89}, 012003 (2014).
\bibitem{17} L.V.~Gribov, E.M.~Levin, M.G.~Ryskin, Phys. Rep. {\bf 100}, 1 (1983);\\
  E.M.~Levin, M.G.~Ryskin, Yu.M.~Shabelsky, A.G.~Shuvaev, Sov. J. Nucl. Phys. {\bf 53}, 657 (1991).
\bibitem{18} S.~Catani, M.~Ciafaloni, F.~Hautmann, Nucl. Phys. B {\bf 366}, 135 (1991);\\
  J.C.~Collins, R.K.~Ellis, Nucl. Phys. B {\bf 360}, 3 (1991).
\bibitem{19} B.~Andersson {\sl et al.} (Small-$x$ Collaboration), Eur. Phys. J. C {\bf 25}, 77 (2002);\\
  J.~Andersen {\sl et al.} (Small-$x$ Collaboration), Eur. Phys. J. C {\bf 35}, 67 (2004);\\
  J.~Andersen {\sl et al.} (Small-$x$ Collaboration), Eur. Phys. J. C {\bf 48}, 53 (2006).
\bibitem{20} A.V.~Lipatov, M.A.~Malyshev, N.P.~Zotov, Phys. Lett. B {\bf 735}, 79 (2014).
\bibitem{21} A.~Szczurek, M.~Luszczak, R.~Maciula, Phys. Rev. D {\bf 90}, 094023 (2014).
\bibitem{22} L.N.~Lipatov, M.I.~Vyazovsky, Nucl. Phys. B {\bf 597}, 399 (2001).
\bibitem{23} A.V.~Bogdan, V.S.~Fadin, Nucl. Phys. B {\bf 740}, 36 (2006).
\bibitem{24} L.N.~Lipatov, Nucl. Phys. B {\bf 452}, 369 (1995); Phys. Rept. {\bf 286}, 131 (1997).
\bibitem{25} M.~Hentschinski, A. Sabio Vera, Phys. Rev. D {\bf 85}, 056006 (2012);\\
  M.~Hentschinski, Nucl. Phys. B {\bf 859}, 129 (2012);\\
  G.~Chachamis, M.~Hentschinski, J.D.~Madrigal Martinez, A.~Sabio Vera, Nucl. Phys. B {\bf 861}, 133 (2012).
\bibitem{26} S.P.~Baranov, A.V.~Lipatov, N.P.~Zotov, Phys. Rev. D {\bf 89}, 094025 (2014).
\bibitem{27} V.A.~Saleev, Phys. Rev. D {\bf 80}, 114016 (2009).
\bibitem{28} J.A.M.~Vermaseren, NIKHEF-00-023 (2000).
\bibitem{29} S.P.~Baranov, A.V.~Lipatov, N.P.~Zotov, Phys. Rev. D {\bf 78}, 014025 (2008).
\bibitem{30} M.~Deak, F.~Schwennsen, JHEP {\bf 0809}, 035 (2008).
\bibitem{31} M.~Ciafaloni, Nucl. Phys. B {\bf 296}, 49 (1988);\\
  S.~Catani, F.~Fiorani, G.~Marchesini, Phys. Lett. B {\bf 234}, 339 (1990);\\
  S.~Catani, F.~Fiorani, G.~Marchesini, Nucl. Phys. B {\bf 336}, 18 (1990);\\
  G.~Marchesini, Nucl. Phys. B {\bf 445}, 49 (1995). 
\bibitem{32} H.~Jung, arXiv:hep-ph/0411287.
\bibitem{33} M.~Deak, H.~Jung, K.~Kutak, arXiv:0807.2403 [hep-ph].
\bibitem{34} F.~Hautmann, M.~Hentschinski, H.~Jung, Nucl. Phys. B {\bf 865} 54, (2012).
\bibitem{35} S.~Catani, F.~Hautmann, Nucl. Phys. B {\bf 427}, 475 (1994); Phys. Lett. B {\bf 315}, 157 (1993).
\bibitem{36} G.~Curci, W.~Furmanski, R.~Petronzio, Nucl. Phys. B {\bf 175}, 27 (1980).
\bibitem{37} F.~Hautmann, M.~Hentschinski, H.~Jung, arXiv:1207.6420 [hep-ph].
\bibitem{38} A.D.~Martin, M.G.~Ryskin, G.~Watt, Phys. Rev. D {\bf 70}, 014012 (2004); Eur. Phys. J. C {\bf 31}, 73 (2003).
\bibitem{39} A.~Kulesza, W.J.~Stirling, Nucl. Phys. B {\bf 555}, 279 (1999).
\bibitem{40} G.P.~Lepage, J. Comput. Phys. {\bf 27}, 192 (1978).
\bibitem{41} H.~Jung, G.P.~Salam, Eur. Phys. J. C {\bf 19}, 351 (2001);\\
  H.~Jung {et al.}, Eur. Phys. J. C {\bf 70}, 1237 (2010).

\end{thebibliography}
\end{document}